\documentclass[twocolumn]{article}
\usepackage{amsmath}
\usepackage{xtab}
\usepackage{breqn}

\begin{document}





\title{The impossibility of ``fairness'': a generalized impossibility result
for decisions}

\author{Thomas Miconi}
\date{} 

\maketitle

\begin{abstract}

Various measures can be used to estimate bias or unfairness in a predictor.
Previous work has already established that some of these measures are
incompatible with each other. Here we show that, when groups differ in
prevalence of the predicted event, several intuitive, reasonable measures of
fairness (probability of positive prediction given occurrence or
non-occurrence; probability of occurrence given prediction or
non-prediction; and ratio of predictions over occurrences for each group) are all mutually
exclusive: if \emph{one} of them is equal among groups, \emph{the other two}
must differ. The only exceptions are for perfect, or trivial (always-positive or
always-negative) predictors. As a consequence, any non-perfect, non-trivial
predictor must necessarily be ``unfair'' under two out of three reasonable sets
of criteria. This result readily generalizes to a wide range of well-known
statistical quantities (sensitivity, specificity, false positive rate,
precision, etc.), all of which can be divided into three mutually exclusive groups.
Importantly, The results applies to all predictors, whether algorithmic or
human. We conclude with possible ways to handle this effect when assessing and
designing prediction methods.

\end{abstract}

\section{Introduction}

Suppose we must make predictions about the occurrence of a certain event, or
condition, when the prevalence of this condition differs among different
population groups. Ideally, we would want our predictions to be free of bias or
prejudice. But what does it mean for a prediction to be ``unbiased''? Simply
equalizing the rate of positive predictions among groups is unsatisfactory, if
the groups do differ in actual prevalence for the condition. We want our
predictions to reflect the different base rates across groups, but not to
unjustly flag people due to their group membership.

Several criteria can be used to determine whether a predictor is biased, by
comparing them across groups (in the following, $Y$ is a binary variable indicating whether a certain event of interest actually occurs in reality,  $Pred$ is a binary variable indicating that the event is predicted to occur by the predictor):

\begin{itemize}

\item Equal probability of positive prediction, given actual occurrence (or non-occurrence): $P(Pred / Y)$ and $P(Pred / \neg Y)$ \emph{(Measure 1)}

\item Equal probability of actual occurrence, given positive (or negative) prediction: $P(Y / Pred)$ and $P(Y / \neg Pred)$ \emph{(Measure 2)}

\item Equal ratio between the number of predicted
and actual occurrences: $P(Pred)/P(Y)$ \emph{(Measure 3)}

\end{itemize}

These conditions have been given different names in the literature. The first
one is known as ``equalized odds'' \cite{hardt2016equality}. The second one
refers to equal \emph{predictive value}, and has
also been called ``predictive parity''. We choose to call the third one ``equal
calibration''\footnote{Here we use ``calibration'' to denote the overall ratio
between the
number of positive predictions and the number of actual occurrences in the
entire group. This is different from \cite{kleinberg2016inherent}, who 
use the term ``calibration'' (within each score bin)  to refer to our Measure 2,
i.e. predictive parity.}.

These quantities provide intuitive measures of bias because if they differ significantly between
groups, we might suspect that the predictor is biased. To illustrate with
examples from criminal justice, if actually-guilty defendants from group A have a lower chance of eliciting
a positive prediction of guilt than actually-guilty members of group B, we would
suspect that the predictor is biased in favour of group A. Conversely, if
defendants from group A that are predicted to be guilty have a much
lower chance of being actually guilty than similarly-predicted members of
group B, we would suspect that the predictor is biased against group A.

The third measure is perhaps the most intuitive: suppose that 30\% of group A
are guilty, but 40\% test positive; while 20\% of group B are guilty, but only
10\% of them test positive. The obvious difference in the severity of the test
across groups (i.e. the $P(Pred) / P(Y)$ ratio) might be regarded as prima
facie evidence of bias.

Ideally, we would want all of these measures to be simultaneously equal across
groups. And indeed, if our predictions were absolutely perfect, with zero
errors (i.e. $P(Y/Pred)=P(Pred/Y)=1$ and $P(Y/\neg Pred)=P(Pred/ \neg Y)=0$, for all groups),
this would actually be the case: all these measures would be strictly equal
across all groups, independent of each group's prevalence for the condition\footnote{Incidentally, checking whether a proposed measure of fairness is met by a perfect predictor is a useful, if weak, criterion for judging such measures; note that equal prediction rates across groups fails to meet it.}

Unfortunately, for non-perfect predictors, when prevalence differs among
groups, some of these measures are known to be incompatible.  Kleinberg,
Mullainathan and Raghavan \cite{kleinberg2016inherent}, as well as Chouldechova
\cite{chouldechova2017fair}, showed that equalizing Measure 2 and Measure 1
among groups simultaneously was impossible. 

The purpose of this note is to show that, for non-perfect
predictors and under different prevalences among groups, all three of these criteria for
fairness are mutually exclusive. That is, unless the predictor is infallible or trivial,
equalizing \emph{one} of these three criteria across groups guarantees that
\emph{both} other criteria will differ across groups. Furthermore, this result
generalizes to a broad range of well-known statistical measures, which can be
shown to fall within three mutually exclusive sets.

\section{Proofs}

The proofs are surprisingly simple, requiring nothing more than the basic axioms
of probability and Bayes' Theorem.

We assume that $P(Y)$ differs among groups, and that the predictor $Pred$ is neither infallible (false positives and false negative both occur), nor trivial (always-positive or always-negative). 

Suppose $P(Pred/Y)$ and $P(Pred/\neg Y)$ are equal among groups (Measure 1). What can we say about $P(Pred)/P(Y)$ (Measure 3)?


\begin{dmath}
P(Pred)/P(Y) = \frac{P(Pred \cap Y) + P(Pred \cap \neg Y)}{P(Y)}\\ 
= \frac{P(Pred / Y) P(Y)}{P(Y)} + \frac{P(Pred / \neg Y) P(\neg Y)}{P(Y)} \\
= P(Pred / Y) + P(Pred / \neg Y) \frac{P(\neg Y)}{P(Y) \label{eq1}}
\end{dmath}

$\frac{P(\neg Y)}{P(Y)}$ is a strictly monotonic (decreasing) function of
$P(Y)$. Therefore, if $P(Pred/Y)$ and $P(Pred/\neg Y)$ are equal across groups,
and $P(Y)$ differs across groups, then expression \ref{eq1} will differ across
groups - unless $P(Pred / \neg Y)$ is 0, which would imply no false positives,
contradicting our assumptions. Therefore, equalizing Measure
1 forces a difference in Measure 3 (and, by simple contraposition, vice versa).

From Bayes' theorem, identical non-zero $P(Pred/Y)$ and different $P(Pred)/P(Y)$ immediately implies different $P(Y/Pred)$. Therefore, equalizing Measure 1 also forces a difference in Measure 2.

From the same reasoning that led to expression 1, simply by swapping symbols,
we obtain that if $P(Y/Pred)$ and $P(Y/\neg Pred)$ are equal among
groups (Measure 2), then $P(Y)/P(Pred)$ is a strictly decreasing
function of $P(Pred)$. Now if $P(Y)/P(Pred)$ were equal among
groups, since $P(Y)$ differs among groups, then $P(Pred)$ would also
need to differ among groups - and thus $P(Y)/P(Pred)$ (as a strictly monotonic
function of $P(Pred)$) would also differ among groups, leading to a
contradiction. Thus, equalizing Measure 2 forces a difference in Measure 3, and
again vice versa from contraposition.

Again, from Bayes' theorem, identical $P(Y/ Pred)$ and different $P(Y)/P(Pred)$ immediately implies different $P(Pred/Y)$. Thus, equalizing Measure 2 forces a difference in Measure 1. 

Collectively, these proofs substantiate the argument of this paper: equalizing
any of these three criteria among groups forces both other criteria to differ among groups.

By following the same methods, we can prove that equalizing both components of
Measure 1 forces a difference in \emph{both} components of Measure 2, and vice
versa. However, this extension does not apply to Measure 3: equalizing Measure 3
only ensures that at least one component of either Measure 2 or Measure 1 will
differ among groups, as shown above.

\section{Application to COMPAS data}

A recent study of the COMPAS recidivism prediction system \cite{angwin2016machine} has generated a spike of interest in the fairness of prediction algorithms.
Among other arguments, the authors pointed out that the algorithm produced a
much more false positives, and much fewer false negatives, for Blacks than for
Whites. The makers of the COMPAS system replied that the algorithm produced equal predictive
value (probability of recidivism, given a positive or negative prediction)
across races \cite{brennan2009evaluating}. Several authors have already pointed out that this discrepancy
was unavoidable \cite{chouldechova2017fair,kleinberg2016inherent}, as also shown in
the present note. 

What was less commonly reported, however, is another discrepancy in the COMPAS
data: Blacks were more likely to elicit positive predictions, \emph{in
proportion} to their actual rate of recidivism - that is, the algorithm was
differently calibrated (or ``harsher'') for Blacks than for whites.
Specifically (using the data published by ProPublica), 1901 out of 3695 Blacks
reoffended (51.4\%) while 2174 out of 3695 were classified as ``high risk''
(58.8\%); whereas 966 out of 2454 Whites reoffended (39.3\%), but only 854 out
of 2454 were classified as ``high risk'' (34.8\%). Thus, Whites were more
likely to reoffend than predicted by the White rate of high-risk classification
($p<10^{-5}$), whereas Blacks were significantly \emph{less} likely to reoffend
than predicted by the Black rate of high-risk classification ($p < 10^{-5}$). 

This is exactly what the results above predict: because predictive value
(probability of occurrence given positive or negative prediction) was roughly
equalized across the groups, and because one group had higher prevalence of
actual recidivism, calibration must necessarily differ.

\section{Generalization}

We note that the results above can immediately be extended to a larger set of
well-known statistical quantities. In particular, note that the components of our Measure 2, $P(Y /
Pred)$ and $P(Y / \neg Pred)$, are equal to the \emph{precision}, (also known as
the predictive positive value)  and the
\emph{false omission rate}, respectively. However, precision is exactly one minus the
false discovery rate ($P(Y / Pred) = 1-P(\neg Y / Pred)$); similarly, the false
omission rate equal one minus the negative predictive value ($P(Y / \neg Pred) =
1-P(\neg Y / \neg Pred)$). Thus, equality or difference in our Measure 2 also
extends to these additional quantities (since equal or different $x$ obviously forces equal
or different values of $1-x$).

Similarly, the components of our Measure 1, $P(Pred /
Y)$ and $P(Pred / \neg Y)$, are equal to the \emph{sensitivity} (also known as
recall, or true positive rate) and the \emph{false positive rate}, respectively.
However, sensitivity is one minus the false discovery rate: $P(Pred / Y) = 1 -
P(\neg Pred / Y)$, and the false positive rate is one minus the \emph{specificity}:
$P(Pred / \neg Y) =1 - P(\neg Pred / \neg Y)$.

As a result, the measures described above can actually be divided into three
mutually exclusive sets, such that equalizing one set across groups guarantees
that at least some components of both other sets will be unequal among groups. These sets
of measures are summarized in Table \ref{thetable}

\begin{table*}
\begin{xtabular}{l p{0.42\textwidth} p{0.42\textwidth}}
 & Expressions & Names \\
 \hline
 & & \\
Set 1 & $P(Y / Pred)$, $P(Y / \neg Pred)$, $P(\neg Y / Pred)$, $P(\neg Y / \neg Pred)$
&Precision (=Predictive Positive Value), False Omission
Rate, False Discovery Rate, Negative Predictive Value \\
 &  & \\
Set 2 & $P(Pred / Y)$, $P(Pred / \neg Y)$,  $P(\neg Pred / Y)$, $P(\neg Pred /
\neg Y)$ & Sensitivity (=Recall, True Positive Rate), False Positive
Rate (=Fallout), False Negative Rate, Specificity \\
 & & \\
Set 3 & $P(Pred) / P(Y)$ & Calibration (=Severity, Harshness Ratio)

\end{xtabular}
\caption{Three mutually exclusive sets of statistical criteria for ``fairness''.
If prevalence $P(Y)$ differs across groups (and assuming a fallible,
non-trivial predictor), equalizing \emph{one} set of measures across groups ensures
that some measures in \emph{both} other sets will
differ across groups. Furthermore, equalizing the first two measures in either
set 1 or 2 ensures that \emph{all} measures in both other groups will differ across
groups.}
\label{thetable}
\end{table*}


\section{Discussion}

Since several intuitive measures of fairness are mutually exclusive (when
populations differ in prevalence and the predictor is neither perfect nor
trivial), it follows that any predictor can always be portrayed as biased or
unfair, by choosing a specific measure of fairness. Note that this result
applies to all forms of prediction, whether performed by algorithms or by
humans.

Given the mathematical impossibility of simultaneously fulfilling all the measures of
``fairness'' described above, how can a predictor adjust their behavior? There are multiple
options, which are not mutually exclusive.

\subsection{Choose one measure and stick to it} 

These various measures tend to emphasize different viewpoints and perspectives
about fairness, and depending on the situation, it is possible that one measure could be legitimately preferred to others.

Measure 3 suggests that the predictor should be equally ``harsh'' across groups, ensuring that the proportion of positive predictions to actual events is similar (ideally, equal to one). This is a highly intuitive notion of fairness at the level of the group.

The second criterion may be seen as reflecting the judge's
viewpoint: ``given the output of the test, what is the chance that the event
actually occurs?'' Similarly, the first criterion seems to reflect the
defendant's viewpoint: ``Given that I am actually innocent (or not), what is the chance that the algorithm will flag me?''

We note that this last measure, in particular, seems to be of special interest. The so-called Blackstone formulation: ``It is better that ten guilty persons escape than that one innocent suffer'' can be interpreted as emphasizing $P(Pred/\neg Y)$ as the most important measure. It is also the basis for Hardt's equalized odds criterion \cite{hardt2016equality} (which in the binary case is exactly identical to our Measure 1).

\subsubsection*{Alternative: Mix and match measures to equalize across criteria}

The above results leave open the possibility of equalizing some individual
quantities within each criterion, at the expense of making others differ. For
example, it would be feasible to equalize calibration/severity, at the same
time as precision (=positive predictive value) and true positive rate
(=sensitivity or recall). This would be at the expense of differing false
positive rates (=fallout, false alarm rate) and negative predictive value.

\subsection{Improve the test as much as possible}

As stated in the introduction, a perfect classifier would simultaneously
equalize all these measures across groups. Inspection of the proofs reveals that
the dependence of these measures on actual prevalence results from terms of the
form  $P(Pred / \neg Y) \frac{P(\neg Y)}{P(Y)}$ - i.e. an error term multiplied
by a strictly monotonic function of the prevalence. If the error terms ($P(Pred
/ \neg Y)$, $P(Y / \neg Pred)$, etc.) can be reduced as much as possible, the
discrepancy between the various measures will also decrease. Thus, improving the
accuracy of the predictors will tend to improve  the compatibility between
various measures of fairness.

\subsection{Ignore these measures and reach for a first-principles solution}

Ideally, a fair predictor would
effectively eliminate group membership as an independent influence on decisions. As a theoretical example, one may consider a multiple-regression approach, in which a
regressor for group membership is used while training the predictor, then
discarded at decision time (this is a purely theoretical example; we do not
claim that such a method is actually feasible or desirable in practice). Under some conditions, a method of this
kind would guarantee that group membership does not affect conclusions
even if other variables correlate with group membership. 

However, such a measure would likely lead to worse
predictions (because group membership may itself be a proxy for other
predictive variables not included in the model). In addition, it also fail some
or all of the
measures of fairness described above, and thus would still be open to claims of bias. 

\subsection{Use a detailed cost-benefit approach}

In some circumstances, the costs of the various types of errors (false alarms,
misses, etc.) can be quantified. In this case, assuming sufficient and
reliable data, the objective optimum for each quantity can be effectively
computed. Such cost-benefits analyses are particularly relevant in the medical
domain, as illustrated by the output of the British National Institute for
Health and Care Excellence. However, in other circumstances, the costs and
benefits are either difficult to assess, or lie in a moral plane that is not
amenable to quantification.

To conclude, we stress that the foregoing should in now way distract from the
important goal of dismantling the very real biases and discriminations that affect
existing decision mechanisms. To spell out the obvious: just because any
decision system will mathematically be ``biased'' along some measure, does not
at all imply that any observed bias is simply a mathematical artifact. Thus, the incompatibility between fairness measures should not be used as a cover for obvious injustices. On the contrary, a better
understanding of logical constraints on the outcomes of decision systems can and should
inform, and therefore assist, efforts toward a fairer world.

\bibliography{smallbiblio}{}
\bibliographystyle{plain}
\end{document}